\documentclass[prb,twocolumn,showpacs]{revtex4}

\usepackage{graphicx}
\usepackage{dcolumn}

\newcommand{\beq}{\begin{equation}}
\newcommand{\eeq}{\end{equation}}
\newcommand{\beqa}{\begin{eqnarray}}
\newcommand{\eeqa}{\end{eqnarray}}

\newcommand{\percent}{\%}

\begin{document}

\title{Metallic interface at the boundary between band and Mott insulators}
\author{S. S. Kancharla$^1$ and E. Dagotto$^{1,2}$}
\affiliation{$^1$Materials Science and Technology Division, Oak Ridge National Laboratory, Oak Ridge, TN 32831 \\
$^2$Department of Physics and Astronomy, University of Tennessee, Knoxville, TN 37996}
\begin{abstract}
Motivated by experiments on atomically smooth layers of LaTiO$_3$, a
Mott insulator, sandwiched between layers of SrTiO$_3$, a band
insulator, a simple model for such heterostructures is studied using
quasi one-dimensional lattices and the Lanczos method.  Taking both
the local and long-range Coulomb interactions into account, and
computing the layer dependent local density of states, a metallic
state was found at the interface whose extent strongly depends on the
dielectric constant of the material. We also observed that the
antiferromagnetic correlations in the bulk Mott phase persist into the
metallic region. Our conclusions are in excellent agreement with
recently reported results for this model in the opposite limit of
infinite dimensions, thus providing an alternative tool to study
electronic reconstruction effects in heterostructures.
\end{abstract}
\date{\today}
\maketitle

\section{Introduction}
The physics of nanostructures comprising atomically smooth layers of
correlated electron materials with abrupt interfaces is a rapidly
growing field spurred by experimental advances. Apart from the
technological promise of new devices, these systems are of fundamental
interest for the novel phases that could occur at the interfaces of
bulk systems. Correlated electron systems in the bulk are
characterized by strong interactions and often exhibit a variety of
competing states 
in close proximity within their phase
diagrams. Fabricating dissimilar layers of these materials it has
become possible to achieve the enhancement of properties compared
to the bulk, such as the polarization increase observed in superlattices
made up of ferroelectric BaTiO$_3$ interspersed with paraelectric
CaTiO$_3$/SrTiO$_3$, as compared to pure BaTiO$_3$ with identical
volume\cite{oakridge}. In other experiments, coexistence of
ferromagnetism and singlet superconductivity has been observed in
artificially fabricated superconductor (S) - ferromagnet
(F) structures\cite{buzdin,bergeret}, as well as the transfer of
charge from a ferromagnet to a superconductor\cite{varela}.

In this work, we shall be concerned with the reorganization of charge
and the interfacial state located at the boundary between a Mott
insulator (MI) and a band insulator (BI), in the context of the
experiments performed by Ohtomo {\it et al}~\cite{ohtomo}. These
experiments involve the fabrication of a system with an arbitrary
number of layers $n$ of LaTiO$_3$, a MI, sandwiched in between $m$
layers of SrTiO$_3$, a BI, on either side. The transport properties of
this heterostructure (HS) revealed metallic behavior at the interface
between the two types of insulators. Pioneering theoretical studies of
this system by Okamoto and Millis \cite{okamoto1,okamotonature} used
the Hartree approximation to treat the long range Coulomb interaction
and a Dynamical Mean Field Theory (DMFT) to treat the local Coulomb
repulsion. DMFT incorporates dynamical correlations exactly in the
limit of infinite coordination and reduces the problem of each
infinite layer to that of a single site coupled to a self-consistent
bath. Consequently, the entire problem is mapped onto a one
dimensional chain of independent impurity problems, coupled to each
other via a self-consistency condition incorporating the hopping
between layers, and a Hartree mean-field for the long range
Coulomb interaction.


In this paper, we employ an alternative approach by considering a
simplified one-band model for a finite heterostructure using a quasi 
one-dimensional geometry.
This limit of
one dimension is the complete opposite to that of infinite dimensions
that has been pursued in previous literature for this
model\cite{okamoto1}. Our main goal is to verify the conclusions
of previous investigations, and provide alternatives to
DMFT for the study of HS systems involving correlated electrons.
We treat the local and non-local Coulomb
interactions exactly using the Lanczos method. Since the lattice
parameters for SrTiO$_3$ and LaTiO$_3$ differ by only $\sim$$1.5 \%$ we
neglect effects of strain at the interface and, as in earlier
work\cite{okamoto1,okamotonature}, we focus on the relevant electronic
degrees of freedom within the Ti-band to define the Hamiltonian. We
obtain the charge profile through the HS, along with the position
dependent local density of states (DOS), and find a very good agreement with
previous results~\cite{okamoto1}. In general, we observe that the
strongest variation of the density moving from a MI to a metallic
region, and then onward to a BI, occurs over a span of 3 to 4 unit
cells. 
Furthermore, the
antiferromagnetic (AF) correlations in the bulk Mott phase persist in
the metallic interface. Similar results for the charge profile are
found by extending our system to a 2-leg ladder. 
\begin{figure}[ht]
\begin{center}
\includegraphics[width=7.5cm,angle=-0] {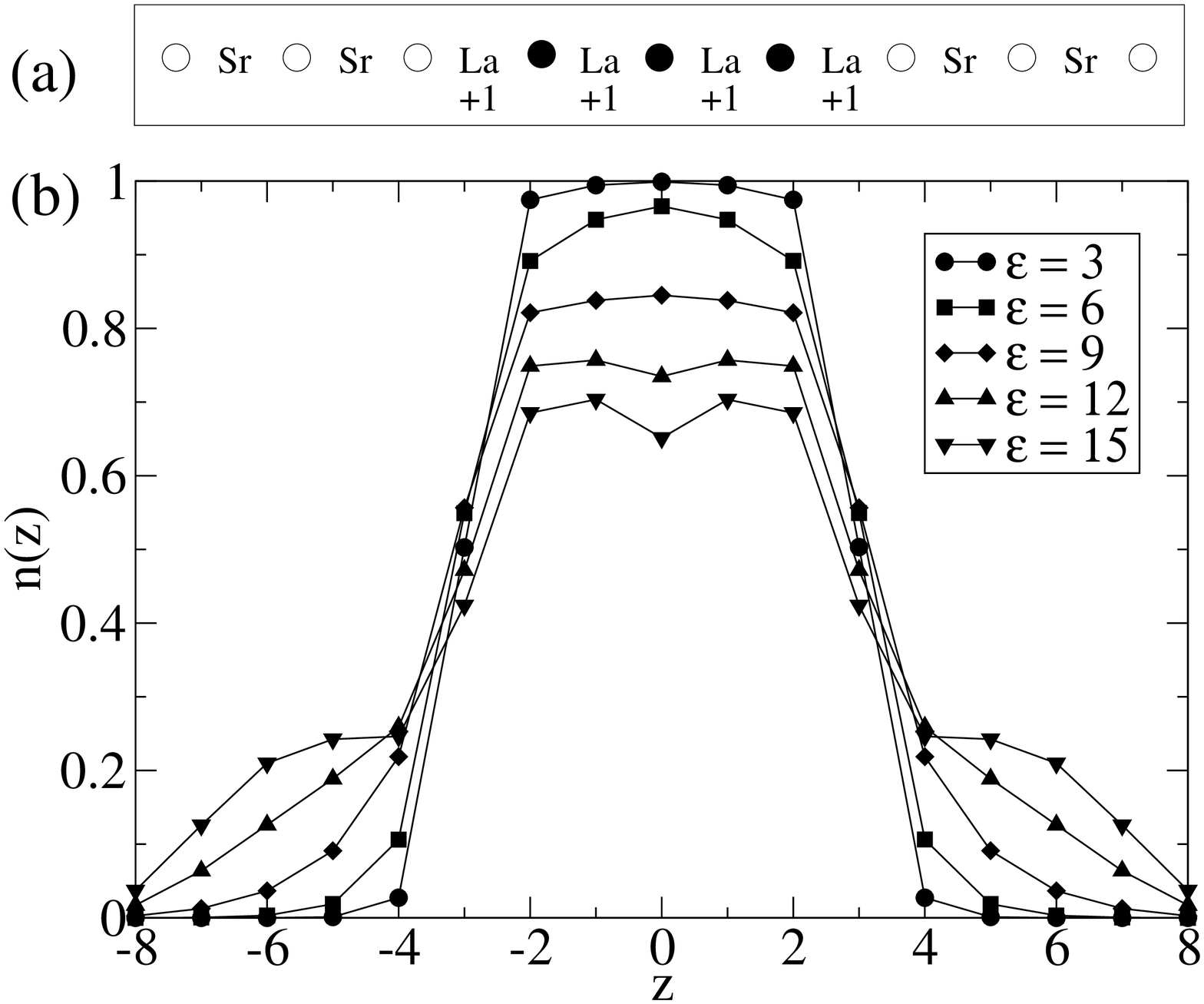}
\end{center}
\caption{(a) Circles denote Ti-sites surrounded by La in the central
Mott region and Sr in the band insulator (b) Charge density across the
16-site heterostructure for $U=20t$ and $\epsilon$ varying from 3 to 15.}
\label{interface}
\end{figure}
\section{Model Hamiltonian}
Consider a HS with $n$ layers of LaTiO$_3$ sandwiched
in between $m$ layers of SrTiO$_3$. A one dimensional analog of this
system is shown in Fig.~\ref{interface}(a). La and Sr carry a bare
valence of +3 and +2 within LaTiO$_3$ and SrTiO$_3$, respectively.  The
condition of charge neutrality requires that substitution of Sr by La
in $n$-layers adds $n$ electrons per unit area of each layer to the
system. The essential dynamics is then governed by
the interplay of the electrons hopping on the Ti-sites, with their mutual
long range repulsion, and the attractive potential produced by the
surplus positive charge of the La ions.

The Hamiltonian for this ($n+2m$)-sized HS in one
dimension can then be written as
\beq
H=H_t + H_U + H_{coul},
\eeq
where
\beq
H_t + H_U = -t \sum_{\langle ij\rangle\sigma ij=1}^{n+2m}
(d^{\dagger}_{i\sigma} d_{j\sigma} + h.c) + U\sum_{i=1}^{n+2m} n_{i\uparrow} n_{i\downarrow},
\eeq
\beq 
H_{coul} = \frac{1}{2}\sum_{i\neq j ij=1}^{n+2m} \frac{e^2
n_{i}n_{j}}{\epsilon|\vec{R_i} -\vec{R_j}|} -\sum_{i=1}^{n+2m}\sum_{j=m+1}^{m+n}\frac{e^2
n_i}{\epsilon|\vec{R_i}-\vec{R_{j}^{'}}|}.
\label{coulomb}
\eeq
Here $H_t$ and $H_U$ represent the nearest neighbor hopping and onsite
Coulomb repulsion through the HS on the Ti-sites. Note that $U$ is
present on all the sites of the HS and not just in the MI region. The
first term in $H_{coul}$ denotes the long range Coulomb repulsion
between the electrons, with $\epsilon$ being the dielectric constant of
the host lattice. The second term represents the long range attractive
potential between electrons on the Ti-sites and the La ions located at
$\vec{R_{j}^{'}} = \vec{R_j} + 1/2 a$, with $a$ being the lattice
constant. The dielectric constant $\epsilon$ is a strong function of
temperature and frequency and also there is considerable uncertainty in the
literature on the values of $U$ for the system. Hence, we investigate
the properties of the above Hamiltonian over a range of parameters,
$8t<U<20t$ and $3<\epsilon<15$ at $T=0$, using the
Lanczos technique\cite{RMP} with open boundary conditions. We set $t = 0.3eV$
and $a=4{\rm \AA}$ for simplicity.
\begin{figure}[htb]
\begin{center}
\includegraphics[width=7.0cm,angle=-0] {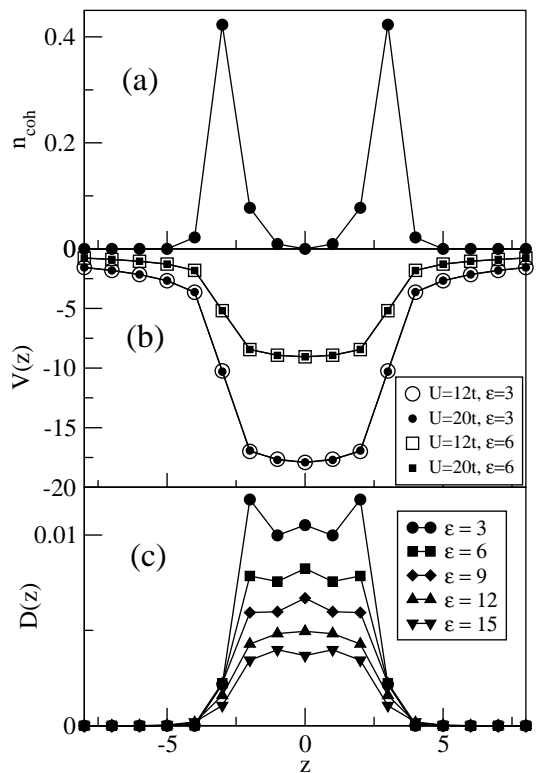}
\end{center}
\caption{(a) Coherent part of the total density, (b) Effective single
particle potential and (c) Double occupancy across the heterostructure for
$U=20t$ and $\epsilon$ varying from 3 to 15.}
\label{mix}
\end{figure}
\section{Results}
Firstly, we present in Fig.~\ref{interface}(b) results for the charge
profile through a 16-layer (i.e. 16-site) HS with $U$=$20t$, as a
function of the dielectric constant $\epsilon$. The structure is made
up of 6 La-ions in the center and corresponds to a net free charge of
6 electrons. We shift the coordinate axis such that the center of the
HS is $z=0$. For $\epsilon=3$, the charge density at the center is
very close to 1 but it rapidly drops over four sites to a very small
number, where the system approaches a BI. The intermediate region
where the density lies between 1 and 0 is expected to be metallic.
The strongest variation of the density moving from a MI to a BI via a
metallic region occurs over 3-4 unit cells.  Increasing $\epsilon$
reduces the confinement of electrons in the central region within the
MI and increases the width of the crossover metallic region. 

Since a metal is characterized by the presence of coherent
quasiparticles, we calculate the coherent fraction\cite{okamoto1} of
the total density $n_{coh}$ along the HS (Fig.~\ref{mix}(a)). This
quantity is defined as an integral over the spectral weight from
$\omega$=$0$ up to the first point in the occupied band where the
spectrum vanishes. We find that $n_{coh}$ has a sharp peak in the
crossover region reaching $\sim$$90\percent$ of the total density,
decreasing to zero in the MI and BI regions. Both the results for the
charge profile and $n_{coh}$ are in excellent agreement with previous
DMFT studies~\cite{okamoto1}. The density profile can be better
understood by constructing an effective one-body potential $V(z)$
which is obtained by substituting the result of the density across the
HS for a given $U$ and $\epsilon$ into $H_{coul}$ and setting
$H_{coul}$ equal to $\sum_z n_z{V(z)}$.  Fig.~\ref{mix}(b) shows
$V(z)$ for $U$=$12t,20t$ and $\epsilon$=$3,6$. It can be observed that
the effective potential that confines the electrons to the MI region
in the center is practically independent of $U$ and only depends on
$\epsilon$. Increasing $\epsilon$ reduces the depth of the potential
well and reduces the slope of the potential in the crossover metallic
region.
\begin{figure}[hbt]
\begin{center}
\includegraphics[width=7.0cm,angle=-0] {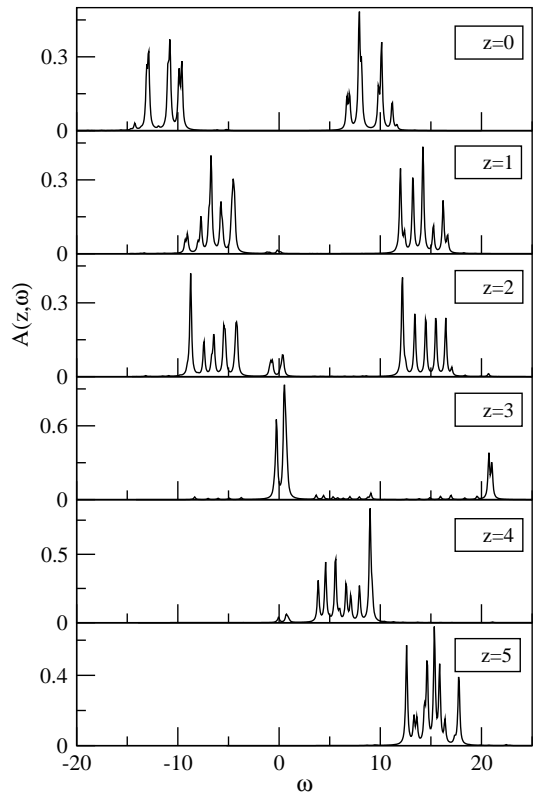}
\end{center}
\caption{Position dependent local density of states along the
16-site heterostructure for $U=20t$ and $\epsilon=3$.}
\label{spectraep3}
\end{figure}

In Fig.~\ref{mix}(c), we plot the double occupancy $D$ along the HS
for several dielectric constants, at a fixed value of $U$=$20t$. Note
that $D$ does not always follow the same behavior as the density
through the HS. For $\epsilon$=$3$, where a clear MI phase prevails in
the central region, $D$ is suppressed at the center ($z$=$0$) relative
to the maximum which occurs in the metallic state at the
interface. Finally $D$ goes to zero as the BI is
approached. Increasing the dielectric constant the double occupancy
shows no dip at the center relative to the crossover region and
essentially follows the behavior of the density, implying that there
is a wider metallic phase that now evolves into a BI with no Mott
phase in the center.
\begin{figure}[htb]
\begin{center}
\includegraphics[width=7.0cm,angle=-0] {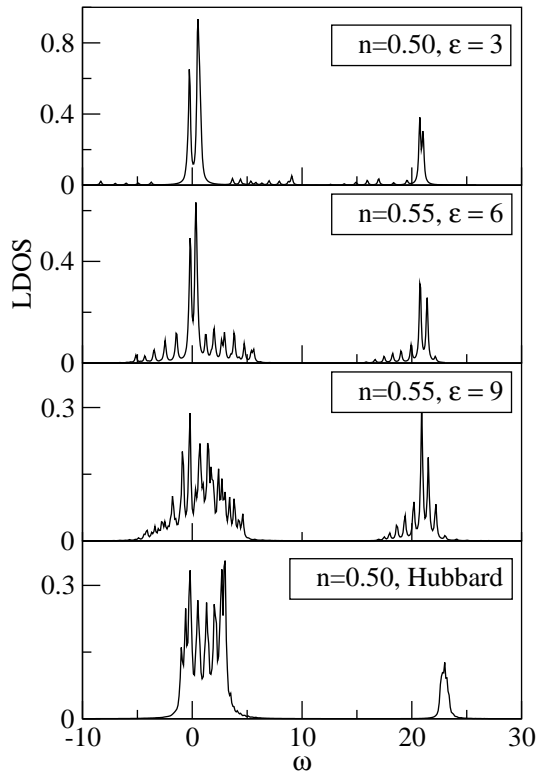}
\end{center}
\caption{Dependence of the local density of states on the dielectric
constant for fixed density, $n\approx0.5$, $U=20t$. Lowest panel shows
the same result for a uniform Hubbard model with periodic boundary
conditions.}
\label{comparison}
\end{figure}

In Fig~\ref{spectraep3}, we show the position dependent local DOS 
along the HS for $U$=$20t$ and $\epsilon$=$3$. At the center,
$z$=$0$, we have a MI with density $n$=$0.999$ reflected by a Mott gap separating
the lower and upper Hubbard bands. Moving to the adjacent
site ($z$=$1$) the density is only slightly lower at $n$=$0.994$ 
but 
a low weight in-gap state appears at the chemical potential.
This metallic phase extends up to site 4, beyond which the
DOS reflects a BI with all the spectral weight above a
finite gap approaching $U$. These results are in excellent agreement with
those obtained earlier for this model using DMFT~\cite{okamoto1}.
\begin{figure}[htb]
\begin{center}
\includegraphics[width=7.0cm,angle=-0] {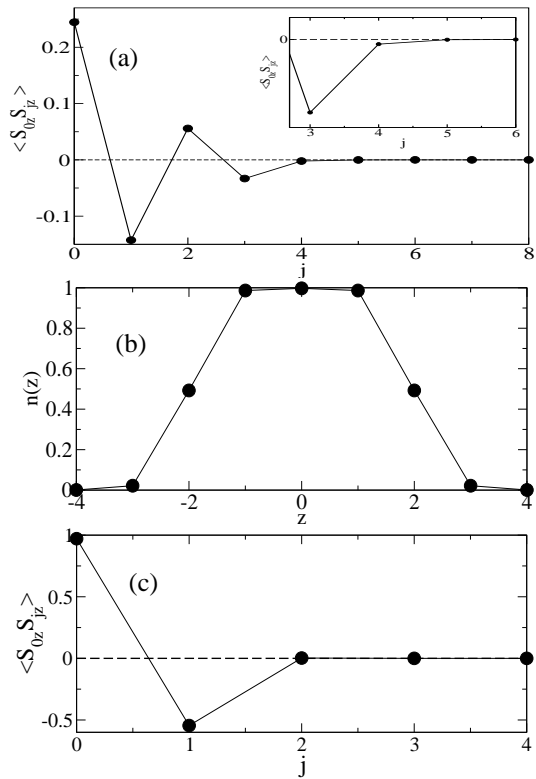}
\end{center}
\caption{(a) Spin-Spin correlation function along the heterostructure.
Inset shows the non-magnetic outer layers (b) Density profile in the
2-leg ladder also shows a metallic interface (c) Spin-Spin correlation
function for the 2-leg ladder along the heterostructure for $U=20t$, $\epsilon=3$.}
\label{spin}
\end{figure}

To clarify the nature of the metallic phase, in Fig.~\ref{comparison}
we plot the local DOS for positions within the HS with comparable
densities ($n$$\approx$$0.5$) but different $\epsilon$, along with the
spectral function for a quarter-filled uniform Hubbard chain of 16
sites under periodic boundary conditions. Both the partially occupied
(lower Hubbard) and unoccupied (upper Hubbard) bands are significantly
widened with the reduction of the dielectric constant and they develop
long tails (although with low weight) as compared to the case with
purely local interactions. For the lowest value $\epsilon$=$3$ most of
the weight in the lower Hubbard band appears in a narrow feature
arising from the onsite Coulomb repulsion, which is well separated
from the tails arising from the long-range part of that interaction.

Next, we present in Fig.~\ref{spin}(a) the spin-spin correlations,
$\langle S_{0z}S_{jz}\rangle$ as a function of distance $j$ from the
central spin at 0. Clearly, AF order is favored in the central region
with density close to 1, as is expected in a MI. As we move toward the
outer layers of the HS, where the density is significantly reduced
from 1 (beyond site 3), there is no alternation of sign in the
spin-spin correlation function signifying a non-magnetic state as we
approach the BI. However, in the intervening metallic interface we
still have a clear signature of AF correlations as nearest-neighbor
correlations (not shown) indicate. To check the robustness of our
results we extend the system perpendicular to the interfacial
direction by considering a 2-leg ladder (a 2$\times$9 system with each
leg containing 2 Sr ions on either side of 4 La ions) with $U$=$20t$
and $\epsilon$=$3$. The density profile shown in Fig.~\ref{spin}(b)
reveals a metallic region in between the MI and BI, as observed before
for the 1D case. In addition, from the spin-spin correlations along
the ladder shown in Fig.~\ref{spin}(c) we find AF correlations in the
central MI. Nearest neighbor AF correlations persist in the metallic
phase (not shown)~\cite{ferro}.

Lastly, we consider a generalization of our model Hamiltonian where
the dielectric constant that governs the attractive interaction of the
ions, $\epsilon_{a}$, differs from the dielectric constant
$\epsilon_r$ for the electron-electron repulsion. In principle, there
is no obvious reason why $\epsilon_{a}=\epsilon_{r}$. Our previous
results simply followed Ref.~6 in these regards, but it is
important to analyse the case $\epsilon_{a}\neq\epsilon_{r}$. The
modified Coulomb term in the Hamiltonian then becomes, \beq H_{coul} =
\frac{1}{2}\sum_{i\neq j ij=1}^{n+2m} \frac{e^2
n_{i}n_{j}}{\epsilon_r|\vec{R_i} -\vec{R_j}|}
-\sum_{i=1}^{n+2m}\sum_{j=m+1}^{m+n}\frac{e^2
n_i}{\epsilon_a|\vec{R_i}-\vec{R_{j}^{'}}|}.  \eeq
\begin{figure}[htb]
\begin{center}
\includegraphics[width=7.0cm,angle=-0] {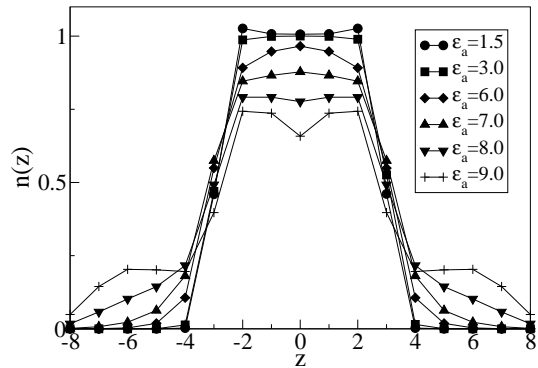}
\end{center}
\caption{Density profile through the HS as a function of $\epsilon_a$
for fixed $\epsilon_r=6$ and $U=20t$.}
\label{gencoul}
\end{figure}
The density profile through the HS for fixed $\epsilon_r=6$ and
varying $\epsilon_a$ is shown in Fig.~\ref{gencoul}. For
$\epsilon_a<\epsilon_r$ we see an increase in the density as we move
away from the central Mott region, so that the density exceeds 1, and
then drops across the metallic interface to approach zero in the BI
region. When $\epsilon_a>\epsilon_r$ we see a suppression of density
below 1 in the central MI and a metallic region which increases in
width at the interface.

\section{Summary and conclusions}
To summarize, motivated by experiments on LaTiO$_3$/SrTiO$_3$
interfaces~\cite{ohtomo}, we have presented an exact diagonalization
study of a simple one-band model for a correlated electron layered
heterostructure in quasi one-dimensional lattices, where a MI is
sandwiched in between a BI. Considering exactly the long range
attractive Coulomb potential of the La ions and the repulsion between
the electrons, we find that the interface supports a metallic state,
in excellent agreement with previous studies of this model using DMFT
and the Hartree techniques~\cite{okamoto1,okamotonature}. 

The strongest
variation of the density across the interface occurs over a region of
3-4 unit cells, whereas the actual width of the metallic region
depends strongly on the dielectric constant of the material. The
charge density depends only weakly on the local Coulomb
interaction. Since the phenomena of interest occur over a very short
spatial range at the interface, we believe that our results are valid
even for HS much larger in the z-direction than the sizes we
considered here. We further check the robustness of our conclusions by
extending the system to a ladder configuration, finding similar
results for the charge profile.  The central region with density close
to 1 shows AF correlations characteristic of a MI which persist into
the metallic phase.

The excellent agreement with DMFT doesn't imply that the self-energy
for our system is relatively momentum independent, but that the
observables such as charge density along the HS are unlikely to be
strongly affected by momentum dependence of the self-energy. Cluster
generalizations of DMFT such as cellular DMFT are likely to provide a
better understanding of this issue. Important directions for future
work using the method discussed here could involve realistic multiband
models for other correlated electron phases including ferromagnetic
and superconducting regimes. Also, the analysis of chain or ladder
systems using density matrix renormalization group or cluster
generalizations of DMFT are possible, although the proper
consideration of long-range Coulomb interactions are likely to pose a
significant challenge to these methods.

S.S.K. is supported by the LDRD program at Oak Ridge National
Laboratory.  E.D. acknowledges support by the NSF Grant
No. DMR-0454504.


\begin{thebibliography}{999}
\bibitem{oakridge} H. N. Lee, H. M. Christen, M. F. Chisholm, C. M. Rouleau, D. Lowndes, Nature (London) \textbf{433}, 395 (2005).
\bibitem{buzdin} A. I. Buzdin, Rev. Mod. Phys., \textbf{77}, 935 (2005).
\bibitem{bergeret} F. S. Bergeret, A. F. Volkov, and K. B. Efetov, Rev. Mod. Phys., \textbf{77}, 1321 (2005).
\bibitem{varela} M. Varela, A. Lupini, V. Pena, Z. Sefrioui, I. Arslan, N. Browning, J. Santamaria, S. J. Pennycook, cond-mat/0508564.
\bibitem{ohtomo} A. Ohtomo, D. A. Muller, J. L. Grazul, and H. Y. Hwang, Nature (London) \textbf{419}, 378 (2002).
\bibitem{okamoto1} S. Okamoto and A. J. Millis, Phys. Rev. B \textbf{70}, 075101 (2004); Phys. Rev. B \textbf{70}, 241104(R) (2004); Phys. Rev. B \textbf{72}, 235108 (2005).
\bibitem{okamotonature} S. Okamoto and A. J. Millis, Nature (London) \textbf{428}, 630 (2004).
\bibitem{RMP} E. Dagotto, Rev. Mod. Phys. {\bf 66}, 763  (1994).
\bibitem{ferro} Indications of ferromagnetism along the rungs of the
ladders were not observed. This is in disagreement with previous
investigations using DMFT~\cite{okamoto1} but certainly a system with
a more extended interface is needed to confirm this result.
\end{thebibliography}
\end{document}